\documentclass[12pt]{article}
\usepackage{latexsym,epsfig,amssymb, amsmath,cite,mcite}
\newcommand{\cM}{{\cal M}}
\newcommand{\cN}{{\cal N}}

\newcommand{\ft}[2]{{\textstyle\frac{#1}{#2}}}

\def\bfone{\relax{\rm 1\kern-.35em 1}}

\newcommand{\be}{\begin{equation}}
\newcommand{\ee}{\end{equation}}
\newcommand{\ben}{\begin{displaymath}}
\newcommand{\een}{\end{displaymath}}
\newcommand{\bea}{\begin{eqnarray}}
\newcommand{\eea}{\end{eqnarray}}

\newcommand{\bean}{\begin{eqnarray*}}
\newcommand{\eean}{\end{eqnarray*}}

\makeatletter \@addtoreset{equation}{section} \makeatother

\begin{document}

\begin{titlepage}

\begin{flushright}
\small UG-09-05\\
\small ENSL-00374319\\
\end{flushright}

\bigskip

\begin{center}

\vskip 2cm

{\LARGE \bf Twin Supergravities} \\[6mm]

{\bf Diederik Roest\,$^1$ and Henning Samtleben\,$^{2}$}\\

\vskip 25pt

$^1$  {\em Centre for Theoretical Physics,\\
University of Groningen, \\
Nijenborgh 4, 9747 AG Groningen, The Netherlands\\
{\small {\tt d.roest@rug.nl}}} \\

\vskip 15pt

$^{2}$ {\em Universit\'e de Lyon, Laboratoire de Physique,\\
Ecole Normale Sup\'erieure de Lyon,\\
46, all\'ee d'Italie, F-69364 Lyon Cedex 07, France\\
{\small {\tt henning.samtleben@ens-lyon.fr}}}

\vskip 0.8cm

\end{center}

\vskip 1cm

\begin{center} {\bf ABSTRACT}\\[3ex]

\begin{minipage}{13cm}
\small

We study the phenomenon that pairs of supergravities can have identical bosonic field content but different fermionic extensions. Such twin theories are classified and shown to originate as truncations of a common theory with more supersymmetry. Moreover, we discuss the possible gaugings and scalar potentials of twin theories. This allows to pinpoint to which extent these structures are determined by the purely bosonic structure of the underlying Kac-Moody algebras and where supersymmetry comes to play its role. 
As an example, we analyze the gaugings of the six-dimensional
${\cal N}=(0,1)$ and ${\cal N}=(2,1)$ theories with identical bosonic sector
and explicitly work out their scalar potentials.
The discrepancy between the potentials finds a natural explanation
within maximal supergravity, in which both theories may be embedded.

\end{minipage}

\end{center}


\vfill

\end{titlepage}

\tableofcontents


\section{Introduction}


Supersymmetry in general poses strong restrictions on the bosonic and fermionic field content of a theory and the possible interactions between them.  A sufficient amount of local supersymmetry even completely fixes the interactions between the different multiplets: the structure of many extended supergravity theories is unique. The only possible deformations are gauge coupling constants and possibly mass parameters. Generically the form of such theories depends on the number of supercharges.

It may therefore come as a surprise that the bosonic sectors of certain specific supergravity theories with different numbers of supersymmetries are in fact identical. In other words, such bosonic sectors allow for different supersymmetric completions. The resulting theories will be referred to as {\em twin} supergravities and are the subject of this paper. As such theories have identical bosons but different fermions (and different supersymmetry), they provide an interesting playground to investigate the role of supersymmetry in various aspects. An example of this concerns the BPS sector of these theories. Black holes that are extremal solutions of both twin theories turn out to be BPS in the one theory and non-BPS in the other \cite{Ferrara:2006yb, Ferrara:2006xx}. That is, the role of BPS and non-BPS sectors are interchanged in the two theories.

In this paper we will study various aspects of twin supergravity theories. First of all, a general classification of twin theories is presented. Secondly, we will discuss how every pair of twin theories originates from a specific supergravity theory, their common {\em parent} theory. Finally, we will also address the possible gaugings and resulting scalar potentials of these theories. Turning on a gauging,
i.e.\ promoting part of its global symmetry group to a local gauge symmetry, will introduce a number of additional terms in the bosonic sector of the theory. On the basis of covariance one can show that most of these terms coincide for the two theories. This is not the case for the scalar potential, however. Indeed, one of the main purposes of this paper is to see whether the effect of such gaugings is different in the scalar potentials; in other words, whether the scalar potentials of gauged twin theories `feel' the different fermionic contents of the two theories. One can in fact argue rather convincingly for both possibilities, i.e.~different or identical scalar potentials. 

An argument supporting the first option is the different amount of supersymmetry of the two theories. In the presence of a gauging, the supersymmetry variations of the fermions acquire additional terms, the so-called fermion shifts that are linear in the gauge coupling constants. The resulting scalar potential takes the form of the difference between the squares of these fermion shifts in order to reconcile the non-abelian deformation with supersymmetry. As the fermionic field content of the two theories is radically different, it is hard to see how their scalar potentials could ever coincide. 
On the other hand, most gauged supergravities are obtained by Kaluza-Klein reductions on particular backgrounds where the scalar potential is generated from reduction of the bosonic part of the higher-dimensional theory on the non-trivial internal geometry. As these bosonic structures coincide for twin theories, gaugings obtained by Kaluza-Klein reduction will exhibit identical scalar potentials. This illustrates the special nature of twin theories, where arguments based on the bosonic part will lead to different expectations than those based on the fermionic part. In this paper we will analyze the structure of gaugings and resolve this paradox. This issue can have a bearing on the possible connection between supergravities and Kac-Moody algebras, as will be discussed in the conclusions, but also on the possible higher-dimensional origin of gaugings.

This paper is organised as follows. The possible twin theories in supergravity are classified in section 2. Subsequently, in section 3 we show how twin theories can be obtained as truncations of particular theories with more supersymmetry. After a discussion of the general pattern, we illustrate these structures with examples in six and four space-time dimensions. Section 4 addresses the possible gaugings of twin theories
using the embedding tensor formalism. We show that these theories admit identical deformations which, however, induce genuinely different scalar potentials. Only for those particular gaugings that can be embedded as gaugings of the common parent theory, the scalar potentials turn out to coincide. We show that these gaugings can be naturally characterized in terms of the quadratic consistency constraints on the embedding tensor of the parent theory.
Section 5 discusses the truncation to twin theories with less supersymmetry where similar structures appear. We conclude in section 6 with some discussion on the importance of these findings for the possible connection between supergravities and Kac-Moody algebras. Finally, appendix A contains some more technical details and explicit formulas of the six-dimensional example.


\section{Classification of Twin Theories}


In this section we will classify the different twin theories, i.e.\ 
pairs of supergravities that have the same bosonic field content and interactions amongst them but which can be supersymmetrically extended with fermions in different ways, leading to different amounts $\cN_\pm$ of supersymmetry. We choose the notation such that 
$\cN_-<\cN_+$. To start with, we focus on the ungauged theories.

\begin{table}[bt]
\begin{center}
\begin{tabular}{||c|rcl||}
\hline ${\cal N}=$ & \multicolumn{3}{c||}{${\cal M}_{\rm scalar}=$} \\ \hline \hline
$1$ &  \multicolumn{3}{c||}{$\rm Riemannian$} \\[1mm] \hline
$2$ &  \multicolumn{3}{c||}{$\rm \mbox{K\"ahler}$} \\[1mm] \hline
$3$ &  \multicolumn{3}{c||}{$\rm Quaternionic$} \\[1mm] \hline
$4$ &  \multicolumn{3}{c||}{${\rm Quaternionic} \otimes {\rm Quaternionic}$} \\[1mm] \hline \hline
$5$ & ${Sp(2,n)}$ & $/$ & ${Sp(2)\otimes Sp(n)}$ \\[1mm]\hline
$6$ & ${SU(4,n)}$ & $/$ & ${SU(4)\otimes SU(n) \times U(1)}$ \\[1mm]\hline
$8$ & ${SO(8,n)}$ & $/$ & $SO(8)\otimes SO(n)$ \\[1mm] \hline
$9$ & ${F_{4(-20)}}$ & $/$ & ${SO(9)}$ \\[1mm] \hline
$10$ & ${E_{6(-14)}}$ & $/$ & ${SO(10)\otimes SO(2)}$ \\[1mm] \hline
$12$ & ${E_{7(-5)}}$ & $/$ & ${SO(12)\otimes SO(3)}$ \\[1mm] \hline
$16$ & ${E_{8(+8)}}$ & $/$ & ${SO(16)}$ \\[1mm] \hline
\end{tabular}
\end{center}
\caption{The scalar manifolds ${\cal M}_{\rm scalar}$ of $D=3$ supergravity theories
with $\cN$ supercharges.} \label{tab:manifolds}
\end{table}

A useful starting point to identify such twin theories is the
classification of $D=3$ supergravity
theories~\cite{deWit:1992up,*deWit:2003ja}. The reason is that in
three dimensions all bosonic fields can be dualized into scalars,
such that the bosonic sectors of the various theories are entirely classified by the
geometry of their scalar manifolds. This allows for a straightforward
comparison. The different scalar manifolds are listen in
table~\ref{tab:manifolds}, where ${\cal N}$ refers to the number of supercharges
(in terms of three-dimensional Majorana spinors; maximal supersymmetry
thus corresponds to ${\cal N}=16$).
The ${\cal N} > 4$ theories
have symmetric scalar manifolds.
For $5 \leq {\cal N} \leq 8$ there is the freedom of including a
number $n$ of matter multiplets, while for $9 \leq {\cal N} \leq 16$
the theories are unique. In contrast, for ${\cal N} \leq 4$ the
scalar manifolds are subject to certain geometric conditions and are
not necessarily symmetric.

The crucial point for the existence of twin theories
is that the geometric conditions on the  ${\cal N}_- \leq 4$ scalar manifolds also happen
to be satisfied by some other theories with a larger ${\cal N}_+$. To start with, all supergravity theories with extended supersymmetry have scalar manifolds that are Riemannian and hence can also be interpreted as an ${\cal N}_-=1$ theory.\footnote{The case of ${\cal M}_{\rm scalar}={E_{8(+8)}}$ with ${\cal N}_- =1$, ${\cal N}_+ = 16$ supersymmetry has been considered
in~\cite{Nishino:2002jv}.} Furthermore, the ${\cal N}_-=3$ theories have a single quaternionic manifold and hence can also be interpreted as $\cN_+=4$ theories with a trivial second factor.

\begin{table}[bt]
\begin{center}
\begin{tabular}{||c|c|rclc|c||}
\hline $\cN_- =$ & $\cN_+=$ & \multicolumn{4}{c|}{$\cM_{\rm scalar}=$} & $D_{\rm max}=$ \\ \hline
\hline
 $2$ & $4$ & $SU(2,n)$ & $/$ & $ SU(2) \times SU(n) \times U(1)$ & ~~(QK) & $4$ \\[1mm] \hline
 $2$ & $6$ & $SU(4,n)$ & $/$ & $ SU(4) \times SU(n) \times U(1)$ & ~~(K) & $4$ \\[1mm] \hline
 $2$ & $8$ & $SO(8,2)$ & $/$ & $ SO(8) \times SO(2)$ & ~~(K) & $4$ \\[1mm] \hline
 $2$ & $10$ & $E_{6(-14)}$ & $/$ & $SO(10) \times SO(2)$ & ~~(K) & $4$ \\[1mm] \hline
 $4$ & $5$ & $Sp(2,1)$ & $/$ & $ Sp(2) \times Sp(1)$ & ~~(Q) & $3$ \\[1mm] \hline
 $4$ & $6$ & $SU(4,2)$ & $/$ & $ SU(4) \times SU(2) \times U(1)$ & ~~(QK) & $4$ \\[1mm] \hline
 $4$ & $8$ & $SO(4,8)$ & $/$ & $ SO(4) \times SO(8)$ & ~~(Q) & $6$ \\[1mm] \hline
 $4$ & $12$ & $E_{7(-5)} $ & $/$ & $SO(12) \times SU(2)$ & ~~(Q) & $6$ \\[1mm] \hline
\end{tabular}
\end{center}
\caption{The twin supergravity theories in three dimensions. $\cM_{\rm scalar}$ denotes the K\"{a}hler, quaternionic or quaternionic-K\"{a}hler scalar
manifold in $D=3$ and $D_{\rm max}$ is the highest dimensions to which
these theories can be uplifted.} \label{tab:twins}
\end{table}

More interesting are those $\cN_+ > 2$ theories which are described by K\"{a}hler (K) manifolds. These can also
be interpreted as ${\cal N}_-=2$ supergravities.  Similarly, there is a number of ${\cal N}_+ > 4$ theories whose bosonic sector is quaternionic (Q). These can therefore be interpreted as an $\cN_- = 4$ theory. An exhaustive list of these latter twin theories is given in table \ref{tab:twins}.

A number of comments is in order. First of all, the
first example in table \ref{tab:twins} with global symmetry group $SU(2,n)$ 
has either two or four supersymmetries. In this case the scalar manifold is both K\"{a}hler
and quaternionic\footnote{A cautionary note on terminology:
following e.g.~\cite{Ferrara:2008de}, quaternionic-K\"{a}hler
manifolds will be understood to have holonomy contained in both
$Sp(d/4) \times Sp(1)$ and in $U(d/2)$, i.e.~they are both
quaternionic and K\"{a}hler. In other conventions a larger set of
manifolds with holonomies contained in $Sp(d/4) \times Sp(1)$ is
referred to quaternionic-K\"{a}hler.} (QK). Moreover, the theory with
$n=4$ has three possible supersymmetric completions with
${\cal N}=2,4$ or $6$ (in addition to the ${\cal N} =1$ and $3$
possibilities that follow from the previous discussion). Finally,
the classification of twin theories in higher dimensions directly follows from that in
three dimensions: all higher-dimensional twins are obtained by dimensional
oxidation of their $D=3$ counterparts. For instance, the twin
theories with highest supersymmetry, i.e.~$\cN_+ = 12$
in table \ref{tab:twins}, can be uplifted to six dimensions. The scalar manifolds of the
higher-dimensional oxidations of this theory are listed in table \ref{tab:example}.

In this paper we will mainly focus on the twin theories that have $\cN_- = 4$ in three dimensions,
as the twin phenomenon is more striking in cases with higher supersymmetry.
Moreover, the structure of the possible gaugings is simpler in these cases,
and there is only a corresponding Kac-Moody algebra for theories with at least eight supercharges.
In the next two sections we will show how these twin theories have an origin in parent
theories with ${\cal N} = {\cal N}_+ + {\cal N}_-$ supersymmetries, and discuss their possible gaugings and scalar potentials. 
We will return to the twin theories with $\cN_- = 2$ in table~\ref{tab:twins} in section 5.

\begin{table}[bt]
\begin{center}
\begin{tabular}{||c||c|c|rcl|}
\hline  $D=$ & ${\cal N}_- =$ & ${\cal N}_+ =$ & \multicolumn{3}{c|}{${\cal M}_{\rm scalar}=$}  \\ \hline
\hline
$6$& $(0,1)$ & $(2,1)$ & $SO(5,1) \times SU(2)$ & $/$ & $SO(5) \times SU(2)$  \\[1mm] \hline
$5$& $2$ & $6$ & $SU^*(6)$ & $/$ & $USp(6)$ \\[1mm] \hline
$4$& $2$ & $6$ & $SO^*(12)$ & $/$ & $U(6)$  \\[1mm] \hline
$3$& $4$ & $12$ & $E_{7(-5)}$ & $/$ & $SO(12) \times SU(2)$ \\[1mm] \hline
\end{tabular}
\end{center}
\caption{The twin theories with highest number of supersymmetry in the different dimensions $3 \leq D \leq 6$.} \label{tab:example}
\end{table}


\section{Parent Theories and Truncations}


In the previous section we discussed the pattern of twin supergravities. 
In particular, in table~\ref{tab:twins}
we identified four different pairs of twin theories with ${\cal N}_- = 4$ and
${\cal N}_+ =5, 6, 8,$ and $12$, respectively. 
In this section we will show how these twin theories can be obtained by truncation
from common parent theories with ${\cal N}={\cal N}_+ + {\cal N}_- 
=9, 10, 12,$ and $16$, respectively. 
The analogous discussion for the four pairs of twins theories with ${\cal N}_- = 2$ and 
${\cal N}_+ = 4, 6, 8, 10,$ respectively, is deferred to section 5.

\subsection{General Structure}

The starting point of this construction is a supergravity theory (to become the parent theory) with ${\cal N}={\cal N}_+ + {\cal N}_-$ supersymmetries, which has a global symmetry group $\hat G$. Two different maximal subgroups of this group will be important in the construction. Firstly, there is the maximal compact subgroup $\hat H \subset \hat G$, which includes the R-symmetry group of the theory. Secondly, we require the existence of a non-compact maximal subgroup of the type $G\times SU(2)$, such that the groups decompose as
\begin{align}
 \hat G~\supset~ G \times SU(2)   \,, \qquad \qquad 
 \hat H ~\supset~H \times SU(2)  \,, 
 \label{decompGH}
\end{align}
where $H$ in turn is the maximal compact subgroup of $G$. The $SU(2)$ factors in \eqref{decompGH} will be crucial in the truncation, as consistency of the truncation will be based on the representations under this group. 
Two different consistent truncations of the parent theory are possible. After decomposing its field content with respect to \eqref{decompGH}, we define the truncations 
\begin{itemize}
\item[{\bf $\mathbb{T}_{+}$\,:}]
  to keep only those fields that satisfy 
 \begin{align}
 (-1)^{F_{SU(2)}} = 1 \,,
 \end{align}
i.e.~that transform in a bosonic representation of the $SU(2)$ factor in \eqref{decompGH}, or
\item[{\bf $\mathbb{T}_{-}$\,:}]
  to keep only those fields that satisfy 
  \begin{align}
  (-1)^{F_{SU(2)}} (-1)^{F_{\rm space-time}} = 1 \,,
 \end{align}
i.e.~space-time bosons that transform in a bosonic representation of the $SU(2)$ factor in \eqref{decompGH} and space-time fermions that transform in a fermionic representation of it.
\end{itemize}
These prescriptions define consistent truncations on group-theoretical grounds. For instance, the Lagrangian contains no terms linear in the fields that are truncated out, as it is a bosonic object with respect to both fermion numbers. Similar arguments hold for the supersymmetry variations (i.e.~the variation of a field that is truncated out consistently vanishes).
Moreover, it is obvious that the two truncations $\mathbb{T}_{\pm}$ give rise to the same bosonic sector but complementary fermionic field content.

Let us consider in detail these truncations 
for the three-dimensional theories collected in table~\ref{tab:manifolds}. 
For the theories with ${\cal N} = 16, 12, 10,$ and $9$, 
the relevant decompositions~\eqref{decompGH} of the 
symmetry groups $\hat G$ are given by
\bea
{\cal N}=16 &:& E_{8(8)} ~\supset~ E_{7(-5)} \times SU(2) 
\;,\nonumber\\[.5ex]
{\cal N}=12 &:& E_{7(-5)} ~\supset~SO(4,8) \times SU(2) 
\;,\nonumber\\[.5ex]
{\cal N}=10 &:& E_{6(-14)} ~\supset~SU(4,2) \times SU(2)  
\;,\nonumber\\[.5ex]
{\cal N}=9 &:&  F_{4(-20)} ~\supset~Sp(2,1) \times SU(2) 
\;,
\label{parentD3}
\eea
respectively. Decomposing their field content, 
and applying the truncation prescription of $\mathbb{T}_{\pm}$, 
defines the following 
reductions of the scalar manifolds
\begin{align}
 \frac{E_{8(8)}}{SO(16)} & \quad \longrightarrow \quad \frac{E_{7(-5)}}{SO(12) \times SO(3)} ~\times~ \frac{SU(2)}{SU(2)} \,, \notag \\[1ex]
 \frac{E_{7(-5)}}{SO(12) \times SO(3)} & \quad \longrightarrow \quad \frac{SO(4,8)}{SO(4) \times SO(8)} ~\times~ \frac{SU(2)}{SU(2)} \,, \notag \\[1ex]
 \frac{E_{6(-14)}}{SO(10) \times SO(2)} & \quad \longrightarrow \quad \frac{SU(4,2)}{SO(6) \times SO(3) \times SO(2)} ~\times~ \frac{SU(2)}{SU(2)} \,, \notag \\[1ex]
 \frac{F_{4(-20)}}{SO(9)} & \quad \longrightarrow \quad \frac{Sp(2,1)}{SO(5) \times SO(3)} ~\times~ \frac{SU(2)}{SU(2)} \,,
 \label{trunc4}
\end{align}
where the scalars transforming in fermionic representations of $SU(2)$
are truncated out.
We recognize as a result the scalar manifolds of 
the ${\cal N}_- = 4$ twin theories
of table~\ref{tab:twins}.
The truncations $\mathbb{T}_{\pm}$  thus define two inequivalent 
truncations of each of the theories of (\ref{parentD3})
which correspond precisely to the pairs of twins identified in the previous section.

To see this, let us consider the fermionic field content of the theories. 
In three dimensions, the R-symmetry group is given by the special orthogonal group acting on the $\cal N$ supersymmetries. 
It is contained in the 
maximal compact subgroup $\hat H$ of the original theory 
and decomposes under the above prescription as 
\begin{align}
 SO({\cal N}) \quad \supset \quad
  SO({\cal N}_+) \times SU(2) \times {SU(2)}  \,,
\end{align} 
where the second $SU(2)$ factor corresponds to the one displayed in
the decomposition of $\hat H$ in~\eqref{decompGH}, \eqref{parentD3}.
The gravitini always transform in the fundamental representation 
of $SO({\cal N})$ and therefore split up according to
\begin{align}
  {\bf \cal N} \quad \longrightarrow \quad ({\bf {\bf \cal N}_+ ,1,1}) \oplus \underline{({\bf 1, 2, 2})} \,,
\end{align}
where we have underlined the fermionic representations with respect to the relevant $SU(2)$.
From this decomposition it follows that the truncation 
$\mathbb{T}_{+}$ preserves ${\cal N}_+$ gravitini and thus gives rise to an ${\cal N}_+$-extended supergravity. In contrast, the truncation $\mathbb{T}_{-}$ retains the other four gravitini and leads to an ${\cal N}_- = 4$ theory. 
Comparing (\ref{trunc4}) to table~\ref{tab:twins} we have thus shown that 
(up to possible deformations and gaugings
to be discussed in the next section)
all pairs of twin theories with ${\cal N}_- = 4$ can be obtained 
by consistent truncation of their respective common parent theory.

Note that we have exhibited explicit $SU(2) / SU(2)$ factors on the right 
hand side of (\ref{trunc4}). Being completely compact these describe no 
scalar degrees of freedom. In the $\cN_- = 4$ theory they have the following interpretation. 
As follows from table 1, three-dimensional ${\cal N}=4$ theories can carry additional hyper-multiplets,
consisting of scalars that span a separate quaternionic manifold. 
The `empty' factor above can be seen to signal the absence of such a manifold, 
and hence of hyperscalars. Due to this, the $SU(2)$ acts as a global 
symmetry in the fermionic sector only.

\begin{figure}[tb]
\begin{center}
 \resizebox{73mm}{!}{\includegraphics{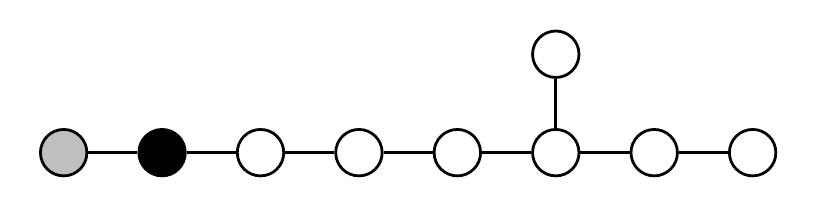}}
\end{center}
\caption{\sl The extended Dynkin diagram of $E_{8(8)}$, where the grey node corresponds to the affine extension and the omission of the black node leads to the relevant semisimple maximal regular subgroup $E_{7(-5)} \times SU(2)$ of $E_{8(8)}$.} \label{fig:E8}
\end{figure}

We further note that the decomposition of the group $\hat G$ into $G \times SU(2)$ has a simple interpretation in terms of the extended Dynkin diagram. Take the Dynkin diagram of $\hat G$ and add one node to obtain the diagram of the (untwisted) affine extension of $\hat G$. The semisimple maximal regular subgroups of $\hat G$ can be obtained by omitting a single node from this extended Dynkin diagram. The subgroups listed in \eqref{parentD3} correspond to the elimination of the node to which the affine node is connected. For the first example in \eqref{parentD3} this is illustrated in figure \ref{fig:E8}.

We have discussed the truncations in detail in three dimensions, but similar decompositions and truncations can be defined in the higher dimensions;
in particular, all higher dimensional examples are obtained by uplift of
table~\ref{tab:twins}.
 For instance, the twin theories with highest amount of supersymmetry, 
 i.e.~$\cN_+ = 12$, can be uplifted to six dimensions. 
The scalar manifolds of the higher-dimensional theories are listed in 
table~\ref{tab:example}. 
In the higher dimensions these twin theories can be obtained similarly by a 
truncation of maximal supergravity. In this case, the relevant 
decompositions of the scalar manifolds are
\begin{align}
 \frac{SO(5,5)}{SO(5) \times SO(5)} & \quad \longrightarrow \quad \frac{SO(5,1)}{SO(5)} \times \frac{SU(2)}{SU(2)} \times \frac{SU(2)}{SU(2)}   \,,  \displaybreak[2] \notag \\[.7ex]
 \frac{E_{6(6)}}{USp(8)} & \quad \longrightarrow \quad \frac{SU^*(6)}{USp(6)} \times \frac{SU(2)}{SU(2)} \,, \displaybreak[2] \notag \\[.7ex]
 \frac{E_{7(7)}}{SU(8)} & \quad \longrightarrow \quad \frac{SO^*(12)}{U(6)} \times \frac{SU(2)}{SU(2)}  \,,
 \label{trunc-example}
\end{align}
in dimensions $D= 6, 5, 4$, respectively.

A crucial ingredient for this higher-dimensional construction to work is that the additional $SU(2)$ factor can be interpreted as part of the R-symmetry group in all these dimensions $3 \leq D \leq 6$ due to the obvious  isomorphisms $SO(3) \sim SU(2) \sim USp(2)$. Furthermore, one can check that in all dimensions these truncations correspond to the maximal regular subgroup that is obtained from the extended Dynkin diagram of $\hat G$ as described above. Finally, the particular truncation of the maximal theory in four dimensions 
was discussed in detail in~\cite{Andrianopoli:2008ea}.

\subsection{Example in Six Dimensions}
\label{subsec:exD6}

It will be useful to illustrate these structures in a concrete example. We will 
consider the ${\cal N}_+ = (2,1)$, ${\cal N}_-=(0,1)$ twin theories in six dimensions. 
These can both be obtained from the maximal six-dimensional supergravity with 
${\cal N} = (2,2)$ supersymmetry. This parent theory has global symmetry group 
$\hat G = SO(5,5)$ and R-symmetry group $\hat H=SO(5) \times SO(5)$.
Its bosonic field content is given by 25 scalars parametrizing the coset space 
$\hat{G}/\hat{H}$, 16 vector fields and 5 antisymmetric tensors. 
The latter combine together with their magnetic duals into the 
vector representation ${\bf 10}$ of $SO(5,5)$.
 Under the decomposition \eqref{decompGH} with $G=SO(5,1) \times SU(2)$ 
we obtain the embeddings
\bea
  SO(5,5) & \supset & SO(5,1) \times SU(2) \times {SU(2)} \,,\nonumber\\
  SO(5) \times SO(5) & \supset & SO(5) \times SU(2) \times {SU(2)} \,.
  \label{decompD6}
\eea
Note that in this case there are in fact two $SU(2)$ factors, of which only the second one will be relevant for the truncation. The bosonic field content of the maximal theory decomposes according to
\begin{alignat}{2}
  \phi: & \quad {\bf 45} \quad & \rightarrow \quad &  ({\bf 1,1,3}) \oplus ({\bf 1,3,1}) \oplus ({\bf 15,1,1}) \oplus \underline{({\bf 6,2,2})} \,, \notag \\
 A_\mu:& \quad {\bf 16}_c \quad & \rightarrow \quad & ({\bf 4,2,1}) \oplus \underline{({\bf 4',1,2})} \,, \notag \\
 B_{\mu\nu}:
 & \quad {\bf 10} \quad & \rightarrow \quad & ({\bf 6,1,1}) \oplus \underline{({\bf 1,2,2})} \,,
 \label{bos6}
\end{alignat}
under $SO(5,1) \times SU(2) \times SU(2) \subset SO(5,5)$. The truncations $\mathbb{T}_\pm$ eliminate the underlined representations which are fermionic
representations under the second $SU(2)$.  The scalars here have been given 
in the adjoint representations. However, as they describe the coset $SO(5,5) / (SO(5) \times SO(5))$, the compact components still need to be modded out. This eliminates the first two components on the right hand side of the first line, while the third component corresponds to the coset $SO(5,1) / SO(5)$ which carries the five 
physical degrees of freedom. The tensor fields split up in five self-dual and five anti-self-dual components, which together transform in the ${\bf 10}$ representation.
After the truncation, five self-dual and one anti-self-dual tensor fields remain.
The truncated theory thus does not straightforwardly admit an action (see however the construction of \cite{PST}), but can be constructed on the level
of the equations of motion along the lines of~\cite{Romans:1986er,*Nishino:1986dc,*Nishino:1997ff,*Riccioni:2001bg}.

The fermions of the parent theory transform in representations of $SO(5) \times SO(5)$, which under (\ref{decompD6}) decompose as
\begin{alignat}{2}
\psi_\mu: & \quad  ({\bf 4,1}) \oplus ({\bf 1,4}) \quad & \rightarrow \quad & ({\bf 4,1,1}) \oplus ({\bf 1,2,1}) \oplus \underline{({\bf 1,1,2})} \,, \notag \\
\chi: & \quad ({\bf 4,5}) \oplus ({\bf 5,4}) \quad & \rightarrow \quad &
  ({\bf 4,1,1}) \oplus \underline{({\bf 4,2,2})} \oplus  \underline{({\bf 5,1,2})} \oplus ({\bf 5,2,1})  \,,
  \label{6dferms}
\end{alignat}
under $SO(5) \times SU(2) \times SU(2)\sim USp(4)\times USp(2)\times USp(2)$. In the fermionic sector, as expected, the truncations $\mathbb{T}_\pm$
give a different result. Under $\mathbb{T}_+$, the underlined components of (\ref{6dferms}) are eliminated. From the gravitini it is clear that this leads to the unique ${\cal N}_+ = (2,1)$ supergravity, studied in~\cite{DAuria:1997cz}. The other truncation $\mathbb{T}_-$ keeps the complementary fermionic representations, i.e.~it keeps only the underlined components of~(\ref{6dferms}). This leads to an ${\cal N}_- = (0,1)$ supergravity coupling eight vector multiplets and five tensor multiplets to minimal supergravity. Like their bosonic truncation, none of these chiral theories admits an action but they can be constructed on the level of the equations of motion.
The entire field content of the ${\cal N}_+ = (2,1)$ comes in singlets under the second $SU(2)$ factor of (\ref{decompD6}), its symmetry group is thus given by $SO(5,1)\times SU(2)$. In contrast, in the ${\cal N}_- = (0,1)$ theory, the second $SU(2)$ factor has a non-trivial action on all fermionic fields.

\subsection{Example in Four Dimensions}
\label{subsec:exD4}

As a second example, let us consider the pair of twin theories in $D=4$
dimensions with ${\cal N}_+=6$, ${\cal N}_-=2$ supersymmetries.
The ungauged theories are obtained by reduction of the previous example,
their common parent theory is again given by maximal supergravity.
Various aspects of this truncation were also discussed recently in~\cite{Andrianopoli:2008ea}.

Maximal supergravity in four dimensions has a global symmetry group $\hat G=E_{7(7)}$, whose maximal compact subgroup is $\hat{H}=SU(8)$. The scalars form the corresponding scalar coset, and can be seen to transform in the adjoint $\bf 133$ of $E_{7(7)}$. Not all of these correspond to physical degrees of freedom. Upon splitting up into $SU(8)$, one finds ${\bf 133} \rightarrow {\bf 63} \oplus {\bf 70}$. The former of these is the adjoint of $SU(8)$ and is projected out due to the coset structure, while the latter representation corresponds to the propagating scalar degrees of freedom. Similarly, the vectors transform in the fundamental $\bf 56$ of $E_{7(7)}$. Under an electric subgroup $SL(8)$ these split into ${\bf 28} \oplus {\bf 28}'$, corresponding to the physical vector fields and their magnetic duals, respectively. The gravitini transform in the ${\bf  8}$ and the dilatini in the ${\bf  56}$ of $SU(8)$.

To define the truncations $\mathbb{T}_\pm$ to the twin theories we will employ the
decomposition (\ref{decompGH}) with $G=SO^*(12)$, leading to
 \begin{align}
  E_{7(7)} \quad \supset \quad SO^*(12) \times {SU(2)} \,, \qquad SU(8) \quad \supset \quad U(6) \times {SU(2)} \,. \label{decompQ}
 \end{align}
Under the former decomposition, 
the $E_{7(7)}$-covariant bosons of maximal supergravity split up into
 \begin{alignat}{2}
 \phi : & \quad & {\bf 133} & \quad \rightarrow \quad ({\bf 66}, {\bf 1}) \oplus ({\bf 1},{\bf 3}) \oplus \underline{({\bf 32}_s,{\bf 2})}  \,, \notag \\
A_{\mu} : & \quad & {\bf 56} & \quad \rightarrow \quad ({\bf 32}_c,{\bf 1}) \oplus \underline{({\bf 12},{\bf 2})} \,.
 \label{bos4}
 \end{alignat}
Similarly, for the fermionic degrees of freedom 
we find
under the latter decomposition of \eqref{decompQ}, 
 \begin{alignat}{2}
 \psi_{\mu} : & \quad & {\bf 8} & \quad \rightarrow \quad ({\bf 6}, {\bf1}) \oplus \underline{({\bf 1},{\bf 2})} \,, \notag \\
 \chi : & \quad & {\bf 56} & \quad \rightarrow \quad ({\bf 20},{\bf 1}) \oplus ({\bf 6},{\bf 1}) \oplus  \underline{({\bf 15},{\bf 2})} \,,
 \label{ferm4}
 \end{alignat}
of $U(6) \times SU(2)$.
The truncations $\mathbb{T}_\pm$ remove the underlined representations
of (\ref{bos4}). I.e.\ the physical vectors of the twin theories together with their magnetic duals transform in the ${\bf 32}_c$ of $SO^*(12)$. 
The scalar fields span the truncated coset space
 \begin{align}
  \frac{E_{7(7)}}{SU(8)} & \quad \longrightarrow \quad \frac{SO^*(12)}{U(6)}  \times \frac{SU(2)}{SU(2)}\,.
 \end{align}
In the fermionic sector, the two truncations $\mathbb{T}_\pm$ 
pick out complementary sets from the parent theory. The $\mathbb{T}_+$ truncation retains six gravitini and 26 dilatini, leading to the $\cN_+ = 6$ theory. In contrast, the $\mathbb{T}_-$ truncations leads to the $\cN_- = 2$ field content of two gravitini and 30 dilatini, required to fill 15 vector multiplets and the supergravity multiplet.
Again, the ${\cal N}_-=2$ theory possesses an additional $SU(2)$ symmetry that
acts exclusively in the fermionic sector. This symmetry is trivial in the $\cN_+ = 6$ theory.


\section{Gaugings and Scalar Potentials}

In this section we study the deformations of pairs of twin theories, i.e.\ 
the possible gaugings of part of their common global symmetry group. 
While these are identical deformations in the original 
bosonic sector of the theory,
they will induce different effects in the fermionic sectors of the twin theories.
As a result, also the scalar potentials induced by the deformation in
the bosonic sector will be found to be genuinely different.

We illustrate the general pattern by means of the two examples 
we have introduced in the previous section.
In particular, we explain how the different potentials of the twin theories
are obtained by truncation from their common parent theory.

\subsection{Example in Six Dimensions}


As a concrete example, let us study the gaugings for the pair of six-dimensional
${\cal N}_+=(2,1)$, ${\cal N}_-=(0,1)$ twin theories 
that we have introduced and discussed in section~\ref{subsec:exD6}.
Following
the general scheme~\cite{Nicolai:2000sc,*deWit:2004nw, *Samtleben:2005bp, deWit:2007mt, Bergshoeff:2007ef}, these gaugings are
encoded in a constant embedding tensor $\theta$ which transforms in the tensor product
of the dual vector field representation with the adjoint representation of the
global symmetry group
\bea
\theta &\subset&
({\bf 4}',{\bf 2},{\bf 1})\otimes \Big( ({\bf 15},{\bf 1},{\bf 1})\oplus ({\bf 1},{\bf 3},{\bf 1}) \Big)
\;.
\label{emb0}
\eea
More explicitly, this tensor projects from the generators $t_\alpha$ of the global
symmetry group onto the generators $X_M$ of the gauge algebra that appear in
the minimal couplings to the vector fields
\bea
D_\mu&=&\partial_\mu - g A_\mu^M\,X_M\;,\qquad
X_M~\equiv~\theta_M{}^\alpha\,t_\alpha
\;. 
\label{cov}
\eea

A closer analysis along the lines of~\cite{deWit:2005hv,*deWit:2008ta}
shows that only particular sub-representations in this tensor product
are allowed in order to define a consistent hierarchy of non-abelian
tensor gauge transformations 
and thus a consistent gauging:
\bea
\theta &:& ({\bf 20},{\bf 2},{\bf 1}) \oplus ({\bf 4}',{\bf 2},{\bf 1}) \;.
\label{emb1}
\eea
The selection of these subrepresentations within (\ref{emb0})
is based on purely bosonic arguments\footnote{Specifically, it is only for this
choice of $\theta$ that the non-abelian gauge algebra induced by (\ref{cov})
can be closed upon using the six antisymmetric tensor fields of (\ref{bos6}),
see~\cite{deWit:2005hv,*deWit:2008ta} for details.}
 and thus independent of the
particular fermionic sector of the theory. 
Nevertheless it turns out that precisely the deformations induced
by parameters (\ref{emb1}) allow for a supersymmetrization with
either ${\cal N}_+=(2,1)$ or ${\cal N}_-=(0,1)$
supersymmetries.
The deformation parameters give rise to fermionic mass terms and  enter 
quadratically  in the scalar potential.
Schematically, in every gauged supergravity 
the fermionic mass terms are of the form
\bea
{\cal L}_{\rm ferm} &=& \bar\psi A \psi + \bar\psi B \chi + \bar\chi C \chi  \;,
\label{Lf}
\eea
where $\psi$ and 
$\chi$ collectively denote the gravitino and spin-1/2 fields, respectively, and
we have suppressed all space-time and internal indices.
The tensors $A$, $B$, and $C$ are obtained by dressing the constant
tensors (\ref{emb1}) with the scalar fields.
The scalar potential in turn takes the schematic form
\bea
{\cal L}_{\rm pot} &=& 
\frac1{2{\cal N}}\,\Big({\rm tr}\,B^2 - \frac{D-1}{2}\,{\rm tr}\,A^2 \Big)
\;,
\label{Lb}
\eea
where ${\cal N}$ is the number of supersymmetries and 
$D$ the space-time dimension.
Even though the 
${\cal N}_+ = (2,1)$ and the ${\cal N}_- = (0,1)$ theory are described by the same 
set (\ref{emb1}) of deformation parameters, their different fermionic field content
implies a different structure of the respective mass tensors $A$, $B$ in (\ref{Lf}),  
and thus a priori a different form of their scalar potentials (\ref{Lb}). In the rest of this section 
we discuss the structure of these terms on the level of representations; 
we give the explicit expressions in appendix~\ref{6dexplicit}.

The ${\cal N}_- = (0,1)$ theory admits an additional class of deformations
in which (part of) the second $SU(2)$ factor is gauged by the vector
fields. The corresponding couplings are described by an additional
component of the embedding tensor
\bea
\lambda &:& ({\bf 4}',{\bf 2},{\bf 3}) \;.
\label{emb2}
\eea
As the second $SU(2)$ acts exclusively on the fermions, these parameters
remain invisible in the bosonic sector except for their quadratic contribution to the
scalar potential. They describe the six-dimensional analogue of the 
local version of the Fayet-Iliopoulos mechanism of four-dimensional
${\cal N} = 1$ supergravity.
Accordingly, we will refer to the parameters $\lambda$ as the 
Fayet-Iliopoulos parameters.

It is instructive to identify the origin of the various deformation
parameters (\ref{emb1}), (\ref{emb2}) within the maximally 
supersymmetric parent theory in six dimensions.
In this theory, the gaugings are described by an embedding tensor
transforming in the ${\bf 144}_c$ of $SO(5,5)$~\cite{Bergshoeff:2007ef}.
Under (\ref{decompD6}) this tensor decomposes according to
\bea
\theta: \quad 
{\bf 144}_c & \rightarrow &
({\bf 20},{\bf 2},{\bf 1}) \oplus ({\bf 4}',{\bf 2},{\bf 1}) \oplus ({\bf 4}',{\bf 2},{\bf 3})
 \oplus  \underline{({\bf 20}',{\bf 1},{\bf 2})} + \notag \\
&& 
 \oplus \underline{({\bf 4},{\bf 1},{\bf 2})} \oplus \underline{({\bf 4},{\bf 3},{\bf 2})}
\;,
\label{emb6}
\eea
and both truncations $\mathbb{T}_\pm$ eliminate the 
underlined components. The remaining representations
are precisely in correspondence with the direct analysis of the twin theories~(\ref{emb1}), (\ref{emb2}).
What is interesting and somewhat unexpected in (\ref{emb6})
is the fact that the truncation from the maximal theory seems to allow for
deformations of the Fayet-Iliopoulos type (\ref{emb2}) even in the 
${\cal N}_+ = (2,1)$ theory where they have not shown up in the direct 
analysis (\ref{emb1}). 
We will see in the following that these are forbidden by an additional
quadratic constraint.

Let us further analyze the structure of deformations of the 
six-dimensional twin theories.
As the fermions in both theories transform under
the compact group $SO(5)\times SU(2)\times SU(2)$, the possible fermionic
mass tensors (\ref{Lf}) are obtained from branching the embedding tensor
(\ref{emb1}), (\ref{emb2})
under this compact group, giving rise to
\bea
({\bf 20},{\bf 2},{\bf 1}) \oplus ({\bf 4}',{\bf 2},{\bf 1}) \oplus ({\bf 4}',{\bf 2},{\bf 3})
&\longrightarrow&
({\bf 16,2,1}) \oplus 2 \cdot ({\bf 4,2,1}) \oplus  ({\bf 4,2,3})
\;.
\label{Ttensor}
\eea
Comparison to the fermionic field content~(\ref{6dferms})
allows to identify the various fermionic mass tensors (\ref{Lf})
in the two theories:
\bea
\begin{tabular}{c|cccc}
${\cal N}_+=(2,1)$  & $\psi_{({\bf 4,1,1})}$ & $\psi_{({\bf 1,2,1})}$ & $\chi_{({\bf 4,1,1})}$ & $\chi_{({\bf 5,2,1})}$\\ \hline
$\psi_{({\bf 4,1,1})}$ & $-$ & $({\bf 4,2,1})$ & $-$ & $({\bf 4 \oplus 16,2,1})$ \\
$\psi_{({\bf 1,2,1})}$ &  & $-$ & $({\bf 4,2,1})$ & $-$  \\
$\chi_{({\bf 4,1,1})}$ & &  &$-$ & $({\bf 4 \oplus 16,2,1})$ \\
$\chi_{({\bf 5,2,1})}$ &&& & $-$ 
\end{tabular}
\label{mass1}
\eea
\bea
\begin{tabular}{c|ccc}
${\cal N}_-=(0,1)$  & $\psi_{({\bf 1,1,2})}$ & $\chi_{({\bf 4,2,2})}$ & $\chi_{({\bf 5,1,2})}$ \\ \hline
$\psi_{({\bf 1,1,2})}$ & $-$ & $({\bf 4,2,1\oplus 3})$ & $-$ \\
$\chi_{({\bf 4,2,2})}$ &  &$-$ & $({\bf 4,2,1\oplus 3}) \oplus ({\bf 16,2,1})$ \\
$\chi_{({\bf 5,1,2})}$ & &   & $-$ 
\end{tabular}
\label{mass2}
\eea
As a result of (\ref{Ttensor}), only two of the various $({\bf 4,2,1})$ blocks are
linearly independent, and all $({\bf 16,2,1})$ coincide. 
The scalar potentials of the two theories thus take the schematic 
expressions
\bea
V_{(2,1)} &=& ({\bf 16,2,1})^2 + ({\bf 4,2,1})^2 + ({\bf 4,2,1})^2 - ({\bf 4,2,1})^2
\;,
\label{pot21}
\\
V_{(0,1)} &=& ({\bf 4,2,1})^2+({\bf 4,2,3})^2
\;,
\label{pot01}
\eea
according to (\ref{Lb}), the squares denoting 
singlets under the compact $SO(5)\times SU(2)\times SU(2)$
(cf.\ (\ref{V21}), (\ref{V01}) for the explicit expressions). 
A priori,  the potentials induced in the twin theories are thus
genuinely different and they furthermore 
explicitly differ from direct truncation 
of the potential of the maximal theory $V_{(2,2)}$. 
In particular, $V_{(0,1)}$ is manifestly positive definite
in contrast to the indefinite potential of the ${\cal N}_+=(2,1)$ theory.
However, as the potentials (\ref{pot21}), (\ref{pot01}) are obtained
from complementary fermionic mass terms (\ref{mass1}), (\ref{mass2})
according to the general relation (\ref{Lb}), it follows that they are 
related by the general identity
\bea
4\,V_{(2,2)} &=& 3\,V_{(2,1)} +V_{(0,1)} 
\;,
\label{relV123}
\eea
where the ${\cal N}=(2,2)$ scalar potential is understood to be truncated to the scalars of the twin theories. Indeed, this relation can be verified for 
the explicit expressions~(\ref{V22red}), (\ref{V21}), and (\ref{V01}).

One of our original questions was the possible discrepancy of the scalar potentials in generic twin theories: do the
deformations that act identically in the bosonic sector 
really give rise to different bosonic scalar potentials,
despite the fact that both potentials (\ref{pot21}), (\ref{pot01})
are obtained by truncating the same ${\cal N}=(2,2)$ potential
of the parent theory to an identical bosonic field content?
In order to answer this question we need to further analyze the
possible consistency constraints on the deformation parameters.
A generic gauging is defined by parameters transforming 
in the representations (\ref{emb1}), (\ref{emb2})
subject to additional quadratic constraints that ensure closure
of the gauge algebra. Some (bosonic) algebra shows that
for the six-dimensional twin theories, these constraints 
which are quadratic in the parameters (\ref{emb1}), (\ref{emb2})
transform according to
\bea
{\cal Q}^{(\theta\theta)}_{\rm constraint} &:& 
2\cdot({\bf 6},{\bf 1},{\bf 1}) \oplus 
({\bf 6},{\bf 3},{\bf 1}) \oplus 
({\bf 10},{\bf 3},{\bf 1}) \oplus 
({\bf 64},{\bf 1},{\bf 1})
\;,
\nonumber\\
{\cal Q}^{(\lambda,\theta)}_{\rm constraint} &:&
({\bf 6},{\bf 1},{\bf 3}) \oplus 
({\bf 6},{\bf 3},{\bf 3}) \oplus 
({\bf 10}',{\bf 1},{\bf 3})
\;,
\label{quadratic12}
\eea
under $SO(5,1)\times SU(2)\times SU(2)$.
Here, ${\cal Q}^{(\theta\theta)}_{\rm constraint}$ denotes the
constraints bilinear in $\theta$ from (\ref{emb1}) which are identical
in the two twin theories, while ${\cal Q}^{(\lambda,\theta)}_{\rm constraint}$
collects the quadratic constraints of the type $\theta\lambda+\lambda\lambda$
that also contain the parameters (\ref{emb2}) and are only non-trivial
in the ${\cal N}_-=(0,1)$ theory.
All these constraints imply various quadratic identities among the
fermionic mass tensors (\ref{Ttensor})
(in particular the so-called supersymmetric
Ward identities). Some of these may thus imply non-trivial identities
among the different forms (\ref{pot21}), (\ref{pot01}) of the scalar potential.
However, an explicit breaking of (\ref{quadratic12}) under the 
compact $SO(5)\times SU(2)\times SU(2)$
shows that the only constraints which are singlets under the compact group
descend from the $({\bf 6},{\bf 1},{\bf 1})$ and thus do not contain 
 the Fayet-Iliopoulos parameters $\lambda\sim({\bf 4}',{\bf 2},{\bf 3})$. As the latter do appear in the $\cN_- = (0,1)$ scalar potential but are absent for $\cN_+ = (2,1)$, 
there is no way to relate the potentials (\ref{pot21}) and (\ref{pot01}) by means of the quadratic constraints
and we conclude that the scalar potentials in the 
${\cal N}_+=(2,1)$ and the ${\cal N}_- = (0,1)$ twin theories
are in general genuinely different.

In order to understand how nevertheless both potentials,
(\ref{pot21}) and (\ref{pot01}), descend from the same 
${\cal N}=(2,2)$ potential upon identical truncation we need to consider the
quadratic consistency constraints analogous to (\ref{quadratic12})
in the parent theory. These are quadratic constraints on the embedding
tensor in the ${\bf 144}_c$ of $SO(5,5)$ which transform
as ${\bf 10} \oplus {\bf 126}_c \oplus {\bf 320}$ under this group~\cite{Bergshoeff:2007ef}.
Breaking these representations down to $SO(5,1)\times SU(2)\times SU(2)$
and comparing to (\ref{quadratic12}) shows that 
there is precisely one additional quadratic constraint
\bea
{\cal Q}^{(\rm max)}_{\rm constraint} &:& 
({\bf 6},{\bf 1},{\bf 1})
\;,
\label{quadratic3}
\eea
that survives the truncations $\mathbb{T}_\pm$.
It gives rise to another quadratic identity among the 
fermionic mass tensors which 
explicitly involves the Fayet-Iliopoulos parameters~$\lambda^2$.
As a result, the discrepancy between the two scalar potentials
\bea
V_{(2,1)}-V_{(0,1)}
\;,
\eea
is a linear combination of the three quadratic consistency
constraints contained in (\ref{quadratic12}) and (\ref{quadratic3}).
Only those deformations of the twin theories whose parameters in 
addition to (\ref{quadratic12}) satisfy the constraints (\ref{quadratic3})
can be embedded as deformations of the maximally supersymmetric parent theory.
For these deformations, the scalar potentials 
(\ref{pot21}) and (\ref{pot01}) coincide despite their seemingly different form.
Generic deformations of the ${\cal N}_+ = (2,1)$ and the ${\cal N}_- = (0,1)$
theory on the other hand will only satisfy~(\ref{quadratic12}) and induce genuinely different scalar potentials.

As we show in appendix~\ref{6dexplicit}, the singlet part within the extra constraint~(\ref{quadratic3})
takes the (schematic) form (cf.\ equation~(\ref{Qmaximal}))
\bea
{\cal Q}^{(\rm max)}_{\rm constraint} & = & 
({\bf 4,2,1})^2+({\bf 4,2,3})^2
\;,
\eea
and in particular admits only real solutions if the Fayet-Iliopoulos
parameters $\lambda$ vanish. The additional constraint therefore 
excludes these additional deformations of the $\cN_- = (0,1)$ theory.

The explicit form of the parameters and constraints discussed 
in this section are given in appendix~\ref{6dexplicit}.
In particular, the explicit form of the potentials~(\ref{pot21}), (\ref{pot01}) 
is given in (\ref{V21}) and (\ref{V01}),
in simplified form in (\ref{Vtwins}).
The additional quadratic constraint (\ref{quadratic3}) from the parent theory
by virtue of which the two potentials can be mapped into each other
is explicitly given in~(\ref{Qmaximal}).


\subsection{Example in Four Dimensions}


We return to our second example: the truncation of $D=4$ maximal supergravity 
to the $\cN_- =2$ or $\cN_+ = 6$ 
twins, described in section~\ref{subsec:exD4}. 
The pattern of their possible gaugings is very analogous to the previous example
and we keep the discussion short.
The embedding tensor which encodes the 
possible gaugings of the maximal theory transforms in the ${\bf 912}$ representation of $E_{7(7)}$~\cite{deWit:2007mt}. Under \eqref{decompQ} it decomposes according to
\bea
\theta: \quad
{\bf 912} &\rightarrow& 
({\bf 352}_s, {\bf 1})  \oplus  ({\bf 32}_c, {\bf 3})
\oplus  \underline{({\bf 220}, {\bf 2})} \oplus  \underline{({\bf 12}, {\bf 2})}
\;.
\eea
of $SO^*(12) \times SU(2)$.  In both truncations the underlined doublet 
components are projected out, and we are left with a singlet and a triplet component which we denote as
 \begin{align}
\theta: & \;\; ({\bf 352}_s,{\bf 1}) \,, \qquad
\lambda: \;\; ({\bf 32}_c, {\bf 3}) \,.
 \label{emb-tens}
 \end{align}
As in the $D=6$ example, these correspond to the deformation parameters
present in both twin theories and 
the Fayet-Iliopoulos parameters of the $\cN_- = 2$ theory, respectively.
The former correspond to a gauging of the $SO^*(12)$ global symmetry
group while the
triplet $\lambda$ corresponds to a gauging of the $SU(2)$ 
R-symmetry
that only affects the $\cN_- = 2$ fermions.

The quadratic constraints to be imposed on these parameters 
for consistency of the gauging of the twin theories transform in the representations
\bea
{\cal Q}^{(\theta\theta)}_{\rm constraint} &:& 
\quad
({\bf 66},{\bf 1})\oplus  ({\bf 462}_s,{\bf 1}) \oplus  ({\bf 2079},{\bf 1})
\;,
\label{QC1}\\
{\cal Q}^{(\lambda,\theta)}_{\rm constraint} &:&
\quad ({\bf 1},{\bf 3}) \oplus  ({\bf 66},{\bf 3}) \oplus 
  ({\bf 495},{\bf 3})
  \;,
\label{QC2}
\eea
analogous to (\ref{quadratic12}).
Only the second set of constraints involves the 
Fayet-Iliopoulos parameters.

On the other hand, gaugings of the maximal theory
satisfy quadratic constraints in the ${\bf 133} \oplus  {\bf 8645}$ of $E_{7(7)}$.
Upon truncation according to $\mathbb{T}_\pm$ 
this gives rise to all of (\ref{QC2}) plus an additional constraint
transforming as
\bea
{\cal Q}^{(\rm max)}_{\rm constraint} &:& 
({\bf 66},{\bf 1})
\;.
\label{QC3}
\eea
I.e.\  those gaugings of the twin theories that descend by truncation from
the maximal supersymmetric parent theory need to satisfy
the additional quadratic constraint~(\ref{QC3}).
It is crucial to note that the $({\bf 66},{\bf 1})$ of (\ref{QC3}) is different from
the the corresponding representation in (\ref{QC1}), it notably contains a contribution
$\Gamma^{MN}_{\alpha\beta} \lambda^{\alpha i} \lambda^{\beta i}$
bilinear in the Fayet-Iliopoulos parameters.

The scalar potentials induced by these deformations
are obtained from (\ref{Lb}) upon dressing (\ref{emb-tens})
with the scalar fields and breaking the representations down
to the compact $U(6)\subset SO^*(12)$. This yields the schematic form
\bea
V_{6} &=& ({\bf 35,1})^2+({\bf 15,1})^2+({\bf 105,1})^2 -({\bf 21,1})^2
\;,
\label{pot6}
\\
V_{2} &=& ({\bf 15,1})^2+({\bf 15,3})^2 - ({\bf 1,3})^2
\;,
\label{pot2}
\eea
see~\cite{Andrianopoli:2008ea} for the explicit expressions.
In particular, the $SU(2)$ triplets $({\bf 1,3})$, $({\bf 15,3})$ descend from
the Fayet-Iliopoulos parameters $\lambda$.
As the only constraint that is bilinear in $\lambda$ and contains
a singlet under the compact group $U(6)\times SU(2)$ is the
additional constraint~(\ref{QC3}), it follows again that the 
two potentials (\ref{pot6}), (\ref{pot2}) coincide only for
those gaugings of the twin theories
that descend from a gauging of the maximal theory.
In this case the gauge parameters are subject to the additional quadratic constraint~(\ref{QC3}) that goes beyond the
quadratic constraints of either of the two twin theories.
It would be interesting to study if, in contrast to the six-dimensional case, (\ref{QC3}) admits solutions with real non-vanishing
Fayet-Iliopoulos parameters $\lambda$.

\section{Truncation to Twins with Less Supersymmetry}

The discussion of the previous two sections has been concerned with twin theories that have $\cN_- = 4$ supersymmetries in three dimensions. 
We will now turn to the twin theories that have ${\cal N}_- = 2$ in three dimensions. As we will see, the situation is similar
but also differs in a number of respects from the discussion in sections 3 and 4. 
We will focus on a specific example, which will highlight all the features of this
case.

Our main example will be the uplift of the case with the
highest amount of supersymmetry to four dimensions, i.e.\ the
fourth row of table~\ref{tab:twins}. In four dimensions, this pair of theories
has $\cN_- = 1$ and $\cN_+ = 5$ supersymmetry, respectively. 
Its parent theory therefore has
$\cN = 6$ and has already been encountered before: it is the four-dimensional
example that arises from the $\mathbb{T}_+$ truncation of maximal
supergravity. Its global symmetry group is $\hat G = SO^*(12)$ and
the maximal compact subgroups is $\hat H = U(6)$. The vectors and
scalars are in the ${\bf 32}_c$ and ${\bf 66}$ of $SO^*(12)$, respectively.
As before, not all of these correspond to propagating degrees of freedom. 
The ${\bf 32}_c$ combines the 16 physical vectors with their magnetic duals
while the physical scalars under $U(6)$ transform according to
${\bf 15}_{+2} \oplus
\overline{\bf 15}_{-2}$, where we have included the $U(1) \subset U(6)$
weights. Furthermore the gravitini are in the ${\bf 6}_{+1}$ and the
dilatini are in the ${\bf 6}_{-5} \oplus {\bf 20}_{+3}$ of $U(6)$.

To define the truncations in this case, one again considers
particular decompositions. The relevant maximal subgroups of $\hat G$ and
$\hat H$ are given by
 \begin{align}
  SO^*(12)\quad \supset \quad SU(5,1) \times U(1)  \,, \qquad U(6) \quad \supset \quad U(5) \times U(1) \,. \label{decompK}
 \end{align}
Under the above, the $SO^*(12)$-representations of vectors and
scalars split up in
 \begin{align}
  A_{\mu} : & \quad {\bf 32}_c \quad \rightarrow \quad {\bf 20}_0 \oplus {\bf 6}_{-1} \oplus \overline{\bf 6}_{1} \,, \notag \\
  \phi : & \quad {\bf 66} \quad \rightarrow \quad {\bf 1}_0 \oplus {\bf 35}_0 \oplus {\bf 15}_1 \oplus \overline{\bf 15}_{-1} & \quad  \,.
 \end{align} 
Similarly, in terms of $U(6)$ and the corresponding decomposition, the fermions split up according to
 \begin{alignat}{2}
  \psi_{\mu} : & \quad {\bf 6}_{+1} \quad & \rightarrow & \quad {\bf 5}_{(1,0)} \oplus {\bf 1}_{(-2,1)} \,, \notag \\
  \chi : & \quad {\bf 6}_{-5} \oplus {\bf 20}_{+3} \quad & \rightarrow & \quad {\bf 5}_{(-2,-1)} \oplus {\bf 1}_{(-5,0)} \oplus {\bf 10}_{(0,-1)} \oplus \overline{\bf 10}_{(3,0)} \,.
 \end{alignat}
Some care needs to be taken in identifying which of the two $U(1)$'s
in the decomposition of $\hat H$ corresponds to the one in the decomposition of $\hat
G$ (the other one comes appears in the decomposition of $G$). In the representations above this one corresponds to the latter
weight.

For the truncation we define the following
conditions with respect to the latter $U(1)$'s in $\hat G$ and $\hat
H$:
\begin{itemize}
\item[{\bf $\mathbb{T}_{+}$\,:}]
  to keep only those fields that have even $U(1)$-weight in \eqref{decompK}, or
\item[{\bf $\mathbb{T}_{-}$\,:}]
  to keep space-time bosons that have even and space-time fermions have odd $U(1)$-weight in \eqref{decompK}.
\end{itemize}
From the decompositions above one can easily infer which field
content these truncations induce. They agree in the
bosonic field content and pick out complementary sets of fermionic
fields. From the gravitini it follows that the $\mathbb{T}_{+}$
truncation leads to the $\cN_+ = 5$ theory, while the
$\mathbb{T}_{-}$ truncation gives rise to $\cN_- = 1$.
In the bosonic sector, the ten physical scalars parametrize the 
coset space $SU(5,1)/U(5)$.

Again it is interesting to consider the effect of the
truncation on the embedding tensor and thus on the possible gaugings. As discussed before, the embedding tensor of the
$\cN=6$ parent theory transforms in the ${\bf 352}_s$ of $SO^*(12)$. Under
the decomposition above this leads to a number of representations
of $SU(5,1) \times U(1)$ with even and odd $U(1)$ charges. Keeping only the former yields
 \begin{align}
  \theta : & \quad {\bf 70}_0 \oplus \overline{\bf 70}_0 \,, \qquad \lambda : \quad {\bf 20}_0 \,, \qquad \xi : \quad {\bf 6}_2 \oplus \overline{\bf 6}_{-2} \,. \label{emb-tensorK}
 \end{align}
The $\theta$ components form the usual embedding tensor of the
$\cN_+ = 5$ theory. The $\lambda$ components can be seen as the
additional Fayet-Iliopoulos parameters for the $\cN_- = 1$ theory
which describe a gauging of the $U(1)$ R-symmetry group.
In this sense they are similar to the additional components we found
in the truncation in section 3. However, the charged
components $\xi$ are a new feature of this truncation. 
To understand the meaning of these additional components we recall that
four-dimensional ${\cal N}=1$ theories admit additional supersymmetric deformations that are not related to any gauging but described by a holomorphic superpotential $W$. In this case, the mass tensors $A$ and $B$ of (\ref{Lf}) are proportional to $W$ and $D_i W$, respectively, where $D_i$ denotes the K\"ahler covariant derivative with respect to the five complex coordinates of the scalar target space $SU(5,1)/SU(5)$. Thus $W$ and $D_i W$ together precisely fill up a complex ${\bf 6}$ of $SU(5,1)$ corresponding to $\xi$. More precisely, we expect the corresponding gaugings of the parent theory to induce, upon the truncation $\mathbb{T}_-$, an ${\cal N}_-=1$ theory with holomorphic superpotential $W=\xi^i {\cal V}_i{}^6(\phi)$, where ${\cal V}_i{}^6(\phi)$ denotes the last column of the $SU(5,1)$ coset representative.
It would be interesting to study these theories in more detail, presumably some quadratic constraints will again put strong restrictions on the possible choices of $\lambda$ and $\xi$.

For the other twins the situation is completely analogous
to the example discussed above. In all cases the parent theory has
${\cal N}_+ + {\cal N}_-$ supersymmetry and its global symmetry
group $\hat G$ has a non-semisimple maximal subgroup of the form
\begin{align}
 \hat G \quad \supset \quad G \times U(1) \,, \qquad \qquad \hat H \quad \supset \quad  H \times U(1) \,, \label{decomp2}
\end{align}
where we also have indicated the decomposition of the R-symmetry
groups. For instance, the relevant decompositions of $\hat G$ and
$\hat H$ in three dimensions are
\begin{alignat}{2}
{\cal N} = 12 : && \quad  \frac{E_{7(-5)}}{SO(12) \times SO(3)} & \quad \rightarrow \quad \frac{E_{6(-14)}}{SO(10) \times SO(2)} \times  \frac{U(1)}{U(1)} \,, \notag \\
{\cal N} = 10 : && \quad  \frac{E_{6(-14)}}{SO(10) \times SO(2)} & \quad \rightarrow \quad \frac{SO(2,8)}{SO(2) \times SO(8)} \times  \frac{U(1)}{U(1)} \,, 
 \label{trunc2} \\
{\cal N} = 8 : && \quad  \frac{SO(8,2n)}{SO(8) \times SO(2n)} &
\quad \rightarrow \quad \frac{SU(4,n)}{SO(6) \times SU(n) \times
SO(2)} \times  \frac{U(1)}{U(1)} \,, \notag \\
{\cal N} = 6 : && \quad  \frac{SU(4,n)}{SO(6) \times SU(n) \times U(1)} & \quad \rightarrow \quad \frac{SU(2,n)}{SO(3) \times SU(n) \times U(1)} \times \frac{SU(2)}{SU(2)} \times  \frac{U(1)}{U(1)} \,. \notag
\end{alignat}
Note that in the last line we find both an empty $SU(2)$ and a
$U(1)$ factor, corresponding to the fact that this theory can be
interpreted with $\cN_+ = 4$ or $\cN_- = 2$, respectively.

In all cases we have checked the truncation of the embedding tensor
yields analogous results to the example discussed above. In
particular, the parent embedding tensor always splits up in the
three types of \eqref{emb-tensorK}: the embedding tensor of the
$\cN_+$ theory, the Fayet-Iliopoulos terms $\lambda$ of the $\cN_-$ theory
plus the additional charged components labelled by $\xi$ that we
suspect to be related to deformations described by a holomorphic superpotential.
The latter generically transform in the fundamental of $G$ and its dual
representation.

\begin{figure}[tb]
\begin{center}
 \resizebox{73mm}{!}{\includegraphics{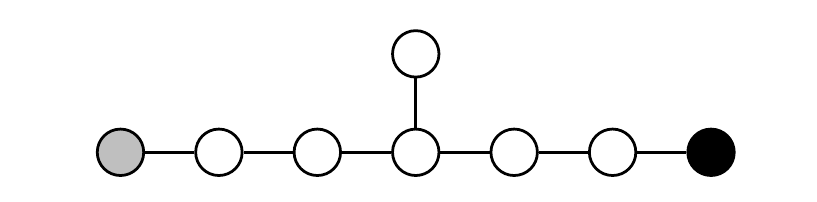}}
\end{center}
\caption{\sl The extended Dynkin diagram of $E_{7(-5)}$, where the grey
node corresponds to the affine extension and the omission of the
black and the grey node leads to the relevant non-semisimple
maximal regular subgroup $E_{6(-14)} \times U(1)$ of $E_{7(-5)}$.} \label{fig:E7}
\end{figure}

Again the subgroups of $\hat G$ defined in \eqref{decomp2}
correspond to maximal regular subgroups, which are non-semisimple
in this case. These can be obtained from the extended Dynkin diagrams by
the deletion of two nodes, after which one has to add an extra
$U(1)$ factor. The example with the highest amount of
supersymmetry in three dimensions, corresponding to the first line
in \eqref{trunc2}, is illustrated in figure \ref{fig:E7}.

\section{Discussion}

We have elucidated a number of aspects of twin supergravities ---
theories with identical bosonic sector but different supersymmetric completion ---
in particular concerning their gaugings and scalar potentials. 
We have given a classification of these theories, and shown that
in general they descend from truncation of a common parent theory.

Two twin theories allow for the same gaugings parametrised by an embedding tensor, while the $\cN_-$-extended theory has the additional possibility to include Fayet-Iliopoulos parameters corresponding to the gauging of symmetries that act exclusively in the fermionic sector. The scalar potentials induced by the gauging in the two theories are genuinely different. They only coincide if the embedding tensor and Fayet-Iliopoulos parameters satisfy an additional quadratic relation that is not required for consistency of the twin theories. The gaugings that satisfy this additional constraint are precisely the ones that can be obtained by truncation from a gauging of the parent theory.
Returning to the discussion in the introduction, this shows in particular that gaugings obtained by dimensional reduction (which by construction exhibit the same scalar potential) do satisfy this extra constraint and can be embedded into the parent theory.
Gaugings of the twin theories that do not satisfy the additional constraint on the other hand, albeit perfectly viable as supersymmetric gaugings of the twin theories, cannot have a higher-dimensional origin.

Although we have only explicitly demonstrated the relation between the two scalar potentials for our two main examples in six and in four dimensions, we have checked that the same structure in terms of representations of quadratic constraints appears in all other twin cases as well. Hence we expect our conclusions to hold for these cases as well. In the six-dimensional example that we discussed in detail, we found that the additional quadratic constraint implies the Fayet-Iliopoulos parameters to vanish. It is not clear whether this result also holds for the other cases.

In addition to the previous results on twin theories that have $\cN_- = 4$ in three dimensions, we have also discussed their $\cN_- = 2$ counterparts. For these theories,  an additional component $\xi$ appears in the truncation of the embedding tensor of the parent theory, that is presumably related to deformations described by a particular holomorphic superpotential. As for the Fayet-Iliopoulos parameters it could be that this component in several cases is eliminated due to the quadratic constraints. We leave this for further study.

As alluded to in the introduction, our results on the gaugings and scalar potentials of twin theories may also be relevant for the connection between supergravities and Kac-Moody algebras. Over the last years, the study of supergravity theories has brought up a number of indications that the structure of these theories is to a large extent determined by the underlying higher-rank Kac-Moody algebras~\cite{West:2001as,*Damour:2002cu}. In particular, many properties that were originally derived from
supersymmetry, such as the field content and the possible deformations (mass parameters and gauge coupling constants) of these theories, were later shown to follow from the
purely bosonic structure of their global symmetry algebras.

In the case of maximal supergravity~\cite{Riccioni:2007au,*Bergshoeff:2007qi,Bergshoeff:2007vb}, the decomposition of the adjoint representation of the very extended algebra $E_{11}$ under suitable subgroups reproduces the field content in $D$ dimensions. Moreover, the non-propagating $(D-1)$-forms correspond to the possible
deformation parameters of the theories, which were found earlier from compatibility with supersymmetry~\cite{Nicolai:2000sc,*deWit:2004nw,*Samtleben:2005bp,deWit:2007mt,Bergshoeff:2007ef}. Finally, the non-propagating $D$-forms correspond to the quadratic constraints on these parameters. Likewise, this information turns out to be encoded in the consistency of the non-abelian gauge algebra of the
higher-rank $p$-forms in a given dimension~\cite{deWit:2005hv,*deWit:2008ta,Bergshoeff:2008qd,*deWit:2008gc,*Bergshoeff:2009ph,*deWit:2009zv}.
A similar picture holds for theories with a lower number of supercharges. Many of these can be associated with a different Kac-Moody algebra, from which the same information can be derived. This was done for the half-maximal supergravities, corresponding to the Kac-Moody extension of $SO(8,8+n)$, in \cite{Bergshoeff:2007vb}. Similarly, the Kac-Moody algebras for the subset of theories with eight supercharges that have symmetric scalar manifolds were discussed in \cite{Gomis:2007gb, *Riccioni:2007hm, *Riccioni:2008jz, Kleinschmidt:2008jj}. A similar analysis can be done for the `exceptional' theories with intermediate amounts of supersymmetry. For theories with less than eight supercharges no  corresponding Kac-Moody algebra is known.

This algebraic correspondence raises the question if the underlying very extended Kac-Moody algebras can encode the information about the full theories, including their dynamics and supersymmetric completions. For instance, to date it is not known if and how the form of the scalar potential is encoded in the very extended Kac-Moody algebras (see \cite{Riccioni:2007ni}, however). The twin supergravities furnish an interesting test ground for this issue, and the analysis in this paper could help to resolve this point. For instance, the discrepancy between the scalar potentials that we have exhibited could find its origin in the different Kac-Moody algebras associated to the two twin theories: for the $\cN_+$-extended theory the associated algebra is the  usual Kac-Moody extension of a simple algebra, while for its $\cN_- = 4$ twin one needs to consider (a quotient of) the Kac-Moody extension of a semisimple algebra \cite{Kleinschmidt:2008jj}. Along a related line of thought, an explicit analysis of the $E_{10}$ $\sigma$-model shows that this yields a positive definite scalar potential, while this is not the case for maximal supergravity in three dimensions~\cite{Bergshoeff:2008xv}. This seems reminiscent of the different scalar potentials \eqref{pot21}, \eqref{pot01} in our six-dimensional example, and may hint at a structure different than maximal supergravity.


\bigskip
\bigskip


\noindent
{\bf Acknowledgements:}
We are grateful to O.~Hohm and M.~Trigiante for helpful discussions.
The work of D.R.\ is supported by a VIDI grant from the Netherlands Organisation for Scientific Research (NWO).
The work of H.S.\ is supported in part by the Agence Nationale de la Recherche (ANR).

\appendix

\section*{Appendix}

\section{Explicit Potentials of the $D=6$ Twin Theories}
\label{6dexplicit}

In this appendix we analyze in detail the example of the
six-dimensional ${\cal N}_+=(2,1)$, ${\cal N}_-=(0,1)$ twin theories
embedded into the maximal ${\cal N}=(2,2)$ theory.
For the latter theory, we use results and notation from~\cite{Bergshoeff:2007ef}.
Its global symmetry group is given by $SO(5,5)$ and the R-symmetry group
by the compact $SO(5)\times SO(5)$.
Fermionic mass tensors in the maximal theory are described in terms
of the embedding tensor, dressed with the scalar fields (\ref{Ttensor}).
This yields two sets of matrices $T_{\alpha\dot\alpha}^a$, $T_{\alpha\dot\alpha}^{\dot{a}}$
related by the linear constraint
\bea
\gamma^a T^a = T^{\dot a} \gamma^{\dot a} \equiv T
\;,
\label{defT}
\eea
with $SO(5)$ gamma matrices $\gamma^a$, $\gamma^{\dot a}$.
Here we use indices $a=1, \dots, 5$ and $\alpha=1, \dots, 4$ for the vector
and the spinor representation, respectively, of the left $SO(5)$ factor of the
R-symmetry group, dotted indices refer to the analogous representations
of the second $SO(5)$ factor. In matrix notation 
we suppress the explicit spinor indices.

Explicitly, the relevant  fermionic mass terms are given by
(see~\cite{Bergshoeff:2007ef} for details)
\bea
{\cal L}_{\rm ferm} &=&
\ft12 {\bar\psi}_{+\mu}\gamma^{\mu\nu}\,( T)\,\psi_{-\nu} 
 + {\bar\psi}_\mu\gamma^\mu \left(T^a-\ft14\gamma^a T\right)\chi^a 
 +{\bar\chi}^{\dot{a}} \left(T^{\dot{a}} +\ft14T\gamma^{\dot{a}} \right) \gamma^\mu\psi_\mu
\nonumber\\
&&+ \dots
\;,
\label{Lfermmax}
\eea
where dots refer to the $(\chi\chi)$ mass terms that are not relevant for the scalar potential.

In accordance with (\ref{Lb}), the scalar potential of the maximal theory is given 
by\footnote{We use the short-hand (but slightly inexact)
notation ${\rm tr} \,T^2 \equiv {\rm tr} \,(T T^{\rm T})$, etc. .
}
\bea
V_{(2,2)} &=&
\tfrac18\,
\Big(
{\rm tr} \left(T^a-\ft14\gamma^a T\right)^2
+{\rm tr} \left(T^{\dot{a}} +\ft14T\gamma^{\dot{a}}\right)^2
- \ft58\, {\rm tr} \,T^2 
\Big)
\nonumber\\
&=&
\tfrac18\,\Big(
{\rm tr} \left(T^a\right)^2
+{\rm tr} \left(T^{\dot{a}}\right)^2
- {\rm tr} \,T^2 
\Big)
\;.
\label{V22}
\eea

Under truncation to
the twin theories, and in agreement with (\ref{Ttensor}), only the following 
components of the $T$ tensor survive:
\bea
T^a &\rightarrow& ({\bf 16,2,1})\oplus ({\bf 4,2,1}) \;,\nonumber\\
T^{\dot a} &\rightarrow& ({\bf 4,2,1})\oplus ({\bf 4,2,1})\oplus ({\bf 4,2,3}) \;,
\eea
where only two of the three $({\bf 4,2,1})$ components are linearly independent.
More explicitly: breaking the second $SO(5)$ factor of the R-symmetry group
according to $SO(5) \rightarrow SU(2)\times SU(2)$ corresponds
to a split of indices
\bea
\dot{\alpha}\rightarrow \{i, \bar\jmath\} \;,\qquad
\dot{a} \rightarrow \{ 0, (i, \bar\jmath) \}
\;,
\eea
with $i, \bar\jmath\in \{1, 2 \}$,
corresponding to the branchings ${\bf 4}\rightarrow ({\bf \frac12,0})\oplus ({\bf 0,\frac12})$
and ${\bf 5}\rightarrow ({\bf 0,0})\oplus ({\bf \frac12,\frac12})$, respectively.
E.g.\ the tensor $T$ from (\ref{defT}) breaks according 
to\footnote{
The only non-trivial input in this branching is the
decomposition of $SO(5)$ $\gamma$-matrices under $SU(2)\times SU(2)$,
for which we use
$\gamma^0_{ij}=\epsilon_{ij},$\, $\gamma^0_{\bar\imath\bar\jmath}=-\epsilon_{\bar\imath\bar\jmath},$
\,$\gamma^{(i, \bar\imath)}_{\,m\bar\jmath}=-\gamma^{(i, \bar\imath)}_{\,\bar\jmath\,m}=
\sqrt{2}\,\delta_m^i\delta^{\bar\imath}_{\bar\jmath}$\,.
}
\bea
T_{\alpha\dot{\alpha}}&\rightarrow&
(T_{\alpha i}, T_{\alpha \bar\imath}) ~=~
\left(
T^0_{\alpha i}-\sqrt{2}\epsilon_{ij} T^{(j,\bar\jmath)}_{\,\alpha\,\, \bar\jmath},
-T^0_{\alpha \bar\imath}-\sqrt{2}\epsilon_{\bar\imath \bar\jmath} T^{(j,\bar\jmath)}_{\,\alpha\,\, j}
\right)
\;.
\eea
Upon truncation to the twin theories, the second component in truncated out,
such that the only non-vanishing component of $T$ is
\bea
T_{\alpha i} &=& T^0_{\alpha i}-T'_{\alpha i} \;,
\qquad 
{\rm with}\;\;
T'_{\alpha i}\equiv \sqrt{2}\,\epsilon_{ij} T^{(j,\bar\jmath)}_{\,\alpha\,\, \bar\jmath}
\;.
\label{blocks}
\eea
We note that in terms of these components
\bea
{\rm tr} \,T^2 &=&
{\rm tr} \left(T^0\right)^2 + {\rm tr} \left(T'\right)^2-2 {\rm tr} \left(T^0 T'{}^{\rm T}\right)
\;,
\nonumber\\
{\rm tr} \left(T^{\dot{a}}\right)^2 &=&
{\rm tr} \left(T^0\right)^2 + {\rm tr} \left(T^{(i,\bar\imath)} \right)^2
\;,
\eea
such that in particular the scalar potential (\ref{V22}) takes the form
\bea
V_{(2,2)} &=&
\tfrac18\,\Big(
{\rm tr} \left(T^a\right)^2
 + {\rm tr} \left(T^{(i,\bar\imath)} \right)^2
 - {\rm tr} \left(T'\right)^2+2\, {\rm tr} \left(T^0 T'{}^{\rm T}\right)
\Big)
\;.
\label{V22red}
\eea
So far, we have just rewritten the maximal gauged ${{\cal N}=(2,2)}$ theory
in terms of the blocks that appear after truncating to the lower ${\cal N}$
theories. In particular, truncation of the potential
to the scalars of (\ref{bos6}) gives rise to the expression (\ref{V22red}).

Let us now study separately the 
gaugings of the two twin theories and their scalar potentials
as derived from their respective supersymmetries.
To this end, we first consider their fermionic mass terms
that are obtained by truncation of (\ref{Lfermmax})
to the complementary fermionic fields (\ref{6dferms}) of the two theories.
Explicitly, this truncation gives rise to
\bea
{\cal N}_+ = (2,1)&:&\;\;
\psi_{({\bf 4,1,1})}:\; \psi^\alpha\,,\quad
\psi_{({\bf 1,2,1})}:\;\psi^i\,,\quad
\chi_{({\bf 4,1,1})}:\;\chi^{0\alpha}\,,\quad
\chi_{({\bf 5,2,1})}:\;\chi^{ai}\;,
\nonumber\\
{\cal N}_- = (0,1)&:&\;\;
\psi_{({\bf 1,1,2})}:\;\psi^{\bar\imath}\,,\quad
\chi_{({\bf 4,2,2})}:\;\chi^{(j,\bar\jmath) \alpha}\,,\quad
\chi_{({\bf 5,1,2})}:\;\chi^{a\bar\imath}\;.
\eea
Accordingly, the fermionic mass terms of the two theories
are obtained from (\ref{Lfermmax}) and yield
\bea
{\cal L}^{(2,1)}_{\rm ferm} &=&
\ft12 {\bar\psi}^\alpha_{\mu}\gamma^{\mu\nu}\,( T_{\alpha i})\,\psi^i_{\nu} 
 + {\bar\psi}^\alpha_\mu\gamma^\mu \left(T^a-\ft14\gamma^a T\right)_{\alpha i}\chi^{a i} 
 +{\bar\chi}^{0 \alpha} \left(T^{0} +\ft14T\gamma^{0} \right)_{\alpha i} \gamma^\mu\psi^i_\mu
\nonumber\\
&&+ \dots
\;,\nonumber\\[2ex]
{\cal L}^{(0,1)}_{\rm ferm} &=&
{\bar\chi}^{(j,\bar\jmath) \alpha} \left(T^{(j,\bar\jmath) } 
+\ft14T\gamma^{(j,\bar\jmath) } \right)_{\alpha  i} \gamma^\mu\psi^i_\mu
~+ \dots
\;,
\eea
respectively.
In accordance with the general form of the scalar potential~(\ref{Lb}), 
${\cal N}_+=(2,1)$ and ${\cal N}_-=(0,1)$ supersymmetry, respectively, thus 
implies that the corresponding scalar potentials are given by
\bea
V_{(2,1)} &=&
\tfrac16\,
\Big(
{\rm tr} \left(T^a-\ft14\gamma^a T\right)^2
+{\rm tr} \left(T^{0} +\ft14T\gamma^{0}\right)^2
- \ft58\, {\rm tr} \,T^2 
\Big)
\nonumber\\
&=&
\tfrac16\,\Big(
{\rm tr} \left(T^a \right)^2
-\ft{1}{4} {\rm tr} \left(T^0\right)^2
-\ft{3}{4} {\rm tr} \left(T'\right)^2
+ 2\, {\rm tr}  \left(T^0 T'{}^{\rm T}\right)
\Big)
\;,
\label{V21}
\eea
and
\bea
V_{(0,1)} &=&
\tfrac12\,\Big(
{\rm tr} \left(T^{(j,\bar\jmath)} +\ft14T\gamma^{(j,\bar\jmath)}\right)^2
\Big)
\nonumber\\
&=&
\tfrac12\,\Big(
{\rm tr} \left(T^{(i,\bar\imath)} \right)^2
+\ft{1}{4} {\rm tr} \left(T^0\right)^2
 - \ft14\,{\rm tr} \left(T'\right)^2
\Big)
\;,
\nonumber\\
&=&
\tfrac12\,\Big(
{\rm tr} \left(\lambda^{(i,\bar\imath)} \right)^2
+\ft{1}{4} {\rm tr} \left(T^0\right)^2
\Big)
\;,
\label{V01}
\eea
respectively.
In the last line, we have split 
 \bea
 T^{(i,\bar\imath)}_{m\,\bar\jmath}=\lambda^{(i,\bar\imath)}_{\alpha\,\bar\jmath}
 -\frac{\sqrt{2}}{4}\delta^{\bar\imath}_{\bar\jmath}\,\epsilon^{ij}\,T'_{\alpha j} \,,
 \label{split}
 \eea
into its trace~$T'$ from (\ref{blocks}) and
a traceless part~$\lambda$ which corresponds to the $({\bf 4,2,3})$
of (\ref{emb2}) and describes the dressed Fayet-Iliopoulos parameters
of the ${\cal N}_- = (0,1)$ theory.
A priori, the potentials induced by the different 
amounts of supersymmetry in the twin theories are
thus genuinely different and also 
explicitly differ from direct truncation 
of the potential of the maximal theory (\ref{V22red}). 
They are however related by the general identity (\ref{relV123})
which indeed can be explicitly verified for (\ref{V22red}), (\ref{V21}), and (\ref{V01}).

In order to understand the possible identification of the various potentials,
we need to consider in more detail the quadratic constraints on the embedding tensor
alluded to in the main text. Any quadratic constraint that gives rise to a singlet
under the compact $SO(5)\times SU(2)\times SU(2)$ gives rise to an identity
that may allow to cast the potentials in formally different though equivalent form.
As we have derived from general arguments above, there are three such constraints
in the maximal theory of which two are also present in the twin theories.
Let us first consider the maximal theory. 
It contains a non-trivial quadratic constraint 
which is a singlet under the compact $SO(5)\times SO(5)$ and reads~\cite{Bergshoeff:2007ef}
\bea
{\rm tr} \left(T^a\right)^2
&=&{\rm tr} \left(T^{\dot{a}}\right)^2
\;.
\label{quad_max}
\eea
In terms of the components~(\ref{blocks}) this implies
\bea
{\rm tr} \left(T^a \right)^2 &=&
{\rm tr} \left(T^0\right)^2+{\rm tr} \left(T^{(i,\bar\imath)} \right)^2 
\;.
\label{qq1}
\eea
Next, there are two quadratic constraints that are vectors under the 
second $SO(5)$ factor and follow from the second equation of 
(3.24) in~\cite{Bergshoeff:2007ef}. These read
\bea
{\rm tr}\, T^a \gamma^{\dot b} \tilde T^a = {\rm tr}\, T^{\dot a}\gamma^{\dot b} \tilde T^{\dot a}
= - {\rm tr}\, T \tilde T^{\dot b}
\;,
\label{quad5}
\eea
and for $\dot b=0$ give rise to two singlet constraints which in terms of the components~(\ref{blocks}) 
take the form
\bea
{\rm tr} \left(T^a \right)^2 &=&
{\rm tr} \left(T^0\right)^2-{\rm tr} \left(T^{(i,\bar\imath)} \right)^2 
\;,
\label{qq2}\\
{\rm tr} \left(T^{(i,\bar\imath)} \right)^2  &=& 
{\rm tr}  \left(T^0 T'{}^{\rm T}\right)
\;.
\label{qq3}
\eea
Recalling from the general discussion in the main text that the quadratic singlet constraints
of the twin theories do not contain the Fayet-Iliopoulos terms, we derive from (\ref{qq1})--(\ref{qq3}) that
(using the split \eqref{split})
\bea
{\cal Q}^{\theta\theta}_{\rm constraint}
&\supset&\quad 
 {\rm tr} \left(T^a \right)^2 - {\rm tr} \left(T^0\right)^2~=~0\;,\quad
{\rm tr}  \left(T^0 T'{}^{\rm T}\right)~=~0\;,\label{Qtwin}\\[1ex]
{\cal Q}^{\rm (max)}_{\rm constraint}
&\supset&\quad 
4\,{\rm tr} \left(\lambda^{(i,\bar\imath)} \right)^2
+ {\rm tr} \left(T'\right)^2 ~=~ 0
\;.
\label{Qmaximal}
\eea
Within the twin theories (i.e.\ making use of ${\cal Q}^{\theta\theta}_{\rm constraint}$), 
we can thus simplify the scalar potentials (\ref{V21}), (\ref{V01}) to
\bea
V_{(2,1)} &=&
\tfrac18\,\Big(
 {\rm tr} \left(T^0\right)^2
- {\rm tr} \left(T'\right)^2
\Big)
\;,
\nonumber\\[1ex]
V_{(0,1)} &=&
\tfrac18\,\Big(
4\,{\rm tr} \left(\lambda^{(i,\bar\imath)} \right)^2
+ {\rm tr} \left(T^0\right)^2
\Big)
\;.
\label{Vtwins}
\eea
This explicitly shows that the twin theories
which admit identical deformations described by an embedding tensor 
satisfying the quadratic constraints (\ref{quadratic12}) 
acquire genuinely different scalar
potentials under these deformations.
Only upon taking into account also the extra quadratic constraint~(\ref{quadratic3}), alias (\ref{Qmaximal}),
that descends from the maximal theory, the two potentials (\ref{Vtwins}) 
coincide.
In this case they both agree with the direct truncation of the potential
of the maximal theory (\ref{V22red}).
In other words, only for those gaugings of the twin theories that can be embedded into 
a gauging of the common parent theory, the two scalar potentials coincide.
Note also that due to the positive definite form of the extra constraint~(\ref{Qmaximal}),
real solutions of this constraint
are only possible for vanishing Fayet-Iliopoulos parameter.


\providecommand{\href}[2]{#2}\begingroup\raggedright\endgroup

\end{document}